\documentclass[aps,pre,twocolumn,showpacs]{revtex4}

\usepackage{graphicx}
\usepackage{epsf} 
\usepackage{dcolumn}
\usepackage{bm}
\usepackage{amssymb}
\usepackage{amsmath}
\usepackage{amsbsy}
\usepackage{graphicx,color,ifthen}
\usepackage{pstricks,pst-node,pst-text,pst-3d,color}
\usepackage[latin1]{inputenc}

\begin{document}

\title[]{Aggregation of chemotactic organisms in a differential flow}
\author{Javier Mu\~noz-Garc\'ia}
\email{javiermunozgarcia@gmail.com}
\homepage[Personal webpage: ]{http://gisc.uc3m.es/~javier}


\affiliation{School of Mathematical Sciences and Complex and Adaptive Systems Laboratory, Systems Biology Ireland, and Grupo Interdisciplinar de Sistemas Complejos (GISC), University
College Dublin, Belfield, Dublin 4, Ireland}

\author{Zolt\'an Neufeld}
\affiliation{School of
Mathematical Sciences and Complex and Adaptive Systems Laboratory, University
College Dublin, Belfield, Dublin 4, Ireland}


\date{\today}

\begin{abstract}
We study the effect of advection on the aggregation and pattern formation in chemotactic systems described by Keller-Segel type models. The evolution of small perturbations is studied analytically in the linear regime complemented by numerical simulations. We show that a uniform differential flow can significantly alter the spatial structure and dynamics of the chemotactic system. The flow leads to the formation of anisotropic aggregates that move following the direction 
of the flow, even when the chemotactic organisms are not directly advected by the flow. Sufficiently strong advection can stop the aggregation and coarsening process that is then restricted to the direction perpendicular to the flow. 
\end{abstract}

\maketitle

\section{Introduction} \label{Introduction}
Directed motion of microorganisms and cells in response to chemical signals - \textit{chemotaxis} -  
plays an important role in a wide range of biological processes including migration of white blood cells, cancer
invasion \cite{anderson:1998}, embryonic development or in locating nutrients by bacteria, algae etc \cite{berg:2000,blackburn:1998}.
In many cases the chemotactic cells not only detect, but also produce chemical signals that may attract
other members of the population. This type of communication based on chemoattractant odors or pheromones 
can control group behavior, aggregation, swarming and collective decisions (quorum sensing) in bacterial colonies \cite{budrene:1991}, slime mold \cite{keller:1970} or insect populations. Often the medium into which the chemical signal is released is not stationary but is a moving fluid (e.g. air or water) while the chemotactic cells or organisms are not transported by the flow as their motility is restricted to crawling on a solid surface. For example, microorganisms may attach to surfaces developing biofilms \cite{donlan:2002} found in natural environments or bioreactors and on a wide variety of surfaces, including living tissues, pipings, and industrial or medical devices. The interface between a surface and an aqueous medium such as water or blood provides an ideal environment for the development of microorganisms. The growth and structure of biofilm communities is a complex process regulated by the properties of the cell surface, diverse characteristics of the medium, type of substratum, and hydrodynamics of the aqueous medium. The influence of hydrodynamics on biofilm structures has been studied recently \cite{neu:1997,stoodley:1999,battin:2003} and was shown that the current velocity affects the structure and dynamics of natural biofilms resulting in different colony shapes \cite{battin:2003}. In  Ref.\ \cite{takayama:1999} is described the use of ``slow'' laminar flows (from 100 $\mu$m s$^{-1}$) to pattern cell culture substrate in capillary systems. Another interesting property which may influence the structure of biofilms is the cell-to-cell signaling or quorum sensing. For example, in Ref.\ \cite{davies:1998} the importance of intercellular molecule signaling on biofilm differentiation is studied.

The response of attached cells to a shear flow and the effects of cell-to-cell signaling on the aggregation have been also studied for a particular slime mold, the \textit{Dictyostelium discoideum}. Using a laminar flow D\'ecav\'e and co-workers have established the critical shear stress for D. discoideum cells on glass \cite{decave:2002} and studied the mechanisms responsible for the induced enhanced motility \cite{decave:2003}. On the oder hand the aggregation of D. discoideum by means of secreting cyclic adenosine monophosphate (cAMP) has been modeled using stochastic and discrete approaches, the continuum descriptions of cell aggregation have been mostly employed and later derived from mechanistic/microscopic descriptions \cite{hillen:2009}. The mathematical properties of these equations is relevant for a broad range models that have been developed to understand the aggregation process in a variety of organisms, pigmentation patterning, neural crest migration, inflammatory response, tumor growth, etc. 

In this work we will study the simultaneous effect of differential advection and cell-to cell signaling on the aggregation and pattern formation of chemotactic biological populations using a model of partial differential equations to describe the evolution of the cell density and the chemical signal concentration. The resulting system is similar to the non-linear chemical reactions studied by Rovinsky and Menzinger involving activator and inhibitor kinetics where a differential flow can induce a pattern forming instability \cite{rovinsky:1992,rovinsky:1994}.

\section{Model} \label{Model}

A well known classical continuum model of chemotaxis at the population level is the Keller-Segel (KS)
model \cite{keller:1970}, that describes the evolution of the density of chemotactic cells, $u(\mathbf{x},t)$, and the chemoattractant concentration, $v(\mathbf{x},t)$, at point $\mathbf{x}$ and time $t$. When the chemical field $v$ is advected by a uniform flow $\mathbf{V}$ and the density field evolves on a fixed substrate we have 
\begin{eqnarray}
&{\partial_t u}= \nabla \cdot \left(D_u \nabla u - \chi(u) \nabla v \right)
\label{Eq.u2}, \\
&{\partial_t v}= D_v \nabla^2 v + f u - s  v - \mathbf{{V}} \cdot \nabla v.
\label{Eq.v2}
\end{eqnarray}
where $D_u$ and $D_v$ are constant diffusivities. The chemoattractant is assumed to be produced proportionally to the local cell density (with a constant of proportionality $f$) while it is degraded with a frequency $s$. Although in Ref. \cite{decave:2003} it was shown that a shear flow increases the cell motility in D. discoideum, since we will consider slower flow velocities than in Ref.\ \cite{decave:2003} (which were of the order of the detachment velocity), in Eq.\ \ref{Eq.u2} we assume that representative values of $D_u$ are closer to those measured in absence of flow such as in Ref.\ \cite{fisher:1989}. Assuming no-flux or periodic boundary conditions the total mass of the biological component is conserved and
can be characterized by the average density. In the original KS model \cite{keller:1970,nanjundiah:1973} $\mathbf{{V}}=0$ and the
chemotactic flux is proportional to the particle density, i.e. $\chi(u)= \chi_0 u$.
Extensions of this model with more general forms
for $\chi(u)$ have also been studied such as the chemotaxis model with prevention of overcrowding introduced in Ref.\ \cite{hillen:2001} where  $\chi(u)=\chi_0 u (1-u/ u_{max})$ with $u_{max}$ the maximum allowed cell density. An important feature of the KS model (observed in biological systems such as slime mold populations) is that it demonstrates an aggregation instability when the total mass of cells is larger then a certain threshold. Properties of the solutions of the KS system 
and its variants have been studied extensively (for recent reviews see Refs. \cite{horstmann:2003,horstmann:2004,hillen:2009}). Interesting analogies between KS type chemotaxis models and nonlinear mean field
Fokker-Planck equations and generalized thermodynamics have been pointed out in \cite{chavanis:2008}.

In order to simplify the analysis of \eqref{Eq.u2} and \eqref{Eq.v2} we introduce non-dimensional variables by rescaling $x' \mapsto (s/D_v)^{1/2} x$, $t' \mapsto st$, $u' \mapsto u_0^{-1} u $, and $v' \mapsto  s(fu_0)^{-1}v$, with $u_0$ the initial mean cell density, resulting the following system:
\begin{eqnarray}
&{\partial_t u}= \nabla \cdot \left(D \nabla u - \chi(u) \nabla v \right)
\label{Eq.u3}, \\
&{\partial_t v}=  \nabla^2 v +  u -   v - \mathbf{{V}} \cdot \nabla v,
\label{Eq.v3}
\end{eqnarray}
where we have defined $D \equiv D_u/D_v$, $\chi(u)'\equiv f/(sD_v)\chi(u)$, $\mathbf{{V}}' \equiv (sD_v)^{-1/2} \mathbf{{V}}$ and omitted primes. In order to estimate the typical spatial and temporal scales in this problem we can use parameter values given in Ref. \cite{nanjundiah:1973} for the chemoattractant cAMP: $D_v \sim 10^{-6}$ cm$^2$ s$^{-1}$ and $s \sim 1$ s. Thus the typical units for the rescaled length, time and velocity are of the order of 0.1 $\mu$m, 1 s and 10 $\mu$m s$^{-1}$ respectively. 

It is important to note that for D. discoideum the critical shear stress for detachment on glass is of the order of $\sigma_{1/2} \approx 2.6$ Pa \cite{decave:2002}. For low Reynolds number when the inertial effects can be neglected the wall shear stress on the adhering cells is proportional to the uniform velocity following $\sigma=6 \eta V/d$, where $\eta$ is the dynamic viscosity of the fluid and $d$ is the distance between the top of the chamber and the substrate. Using $d=0.25$ mm as in Ref.\ \cite{decave:2003} and water at room temperature ($\eta=10^{-3}$ Pa s) we obtain an estimate for the detachment velocity:  $V_{1/2} \approx 10^{5}$ $\mu$m s$^{-1}$. Thus, as we will see below, the flow velocities considered here are smaller than the velocity needed to detach the cells from the substrate and we can assume that the organisms are not advected by the flow, although they are still able to move by crawling on the solid surface as represented by the chemotactic and diffusive terms.


\section{Linear Analysis} \label{Linear}

In order to gain insight into the system we consider the 
stability of the spatially uniform solution. Assuming uniform initial conditions for $u$ and $v$ with $v(t=0)\equiv v_0$ all spatial derivatives vanish in Eqs.\ \eqref{Eq.u3} and \eqref{Eq.v3} and we have the
solutions $u(t)= 1$ and $v(t)=1+(v_0-1) e^{-t}$.
Thus, independently of the initial conditions the concentration tends to 1 for large times. To investigate pattern forming instability in this system we consider the evolution of spatially non-uniform periodic perturbations 
added to the uniform steady state of the form 
$u(\mathbf{x},t) = 1 + \hat{u} \exp \left[ i\mathbf{q \cdot x} +
\omega(\mathbf{q}) t \right]$ and $v(\mathbf{x},t) = 1 + \hat{v} \exp \left[ i\mathbf{q \cdot x} +
\omega(\mathbf{q}) t \right]$, and study whether the perturbation is amplified or damped out in the course of time. 
Here $\mathbf{q}$ is the wave vector of the perturbation, $\omega(\mathbf{q})$ is the corresponding dispersion
relation, and $\hat{u}$ and $\hat{v}$ are the amplitudes of the perturbation at $t=0$. 
Substituting these expressions into \eqref{Eq.u3} and \eqref{Eq.v3}, and
neglecting quadratic terms in $\hat{u}$, $\hat{v}$, we obtain a
linear system of equations where non-trivial solutions only exist if the determinant of
the coefficient matrix is equal to zero. In contrast to previous chemotactic models where advection is not considered (see for example \cite{chavanis:2006} for a linear stability study of an inertial model generalizing the KS model), the resulting quadratic equation for the dispersion relation has complex coefficients and reads
\begin{equation}\label{omega}
    \omega^2+\omega(a+ib)+(c+id)=0.
\end{equation}
The coefficients $a$, $b$, $c$, and $d$ are functions of parameters and
wave-vector components yielding
\begin{align}
    a&=(D+1){q}^2+1 \label {W-a} \\
    b&=\mathbf{{V}} \cdot \mathbf{q} \label {W-b} \\
    c&=D {q}^4 + \left[ D - \chi(1) \right] {q}^2 \label {W-c} \\
    d&= D q^2 \mathbf{{V}} \cdot \mathbf{q}. \label {W-d}
\end{align}
The real and imaginary parts corresponding to the two complex solutions are (see for example page 95 of \cite{mostowski:1964})
\begin{align}
   {\rm Re}(\omega^\pm) &=  -\frac{a}{2} \pm
   \frac{1}{2\sqrt{2}} \{[(a^2-b^2-4c)^2+(2ab-4d)^2]^{1/2} \nonumber \\
&+a^2-b^2-4c \}^{1/2} ,\label{ReW+-}\\
  {\rm Im}(\omega^\pm) &=  -\frac{b}{2} \pm \frac{{\rm sgn}(2ab-4d)}{2\sqrt{2}}
   \{[(a^2-b^2-4c)^2  \nonumber \\
  & +(2ab-4d)^2]^{1/2} -a^2+b^2+4c \}^{1/2}. \label{ImW+-}
\end{align}
The real part of $\omega$ gives the growth or decay rate of the perturbation
amplitude. In particular, the mode corresponding to the maximum of ${\rm Re}(\omega)$, which we denote by $\mathbf{q}^{\it l}$, determines the characteristic wavelength of the pattern in the linear regime, while the imaginary part describes its propagation in space. The velocity $\mathbf{V^{\it l}}$ at which the instability travels across the substrate, corresponding to the \textit{phase velocity} of the mode $\mathbf{q}^{\it l}$, satisfies the relationship \cite{mattheij:2005} 
\begin{equation}\label{Vl}
\mathbf{V^{\it l}} \cdot \mathbf{q}^{\it l} = - {\rm Im}\left[\omega( \mathbf{q}^{\it l})\right]. 
\end{equation}

The negative branch of the dispersion relation is unconditionally stable, i.e.  ${\rm Re}(\omega^-)<0$, but
the positive branch may produce non-trivial dynamics and pattern formation for a certain range of wavenumbers.
Some insight into the behavior of the system can be obtained
by investigating the limiting cases of small and large wavelengths. 
For large values of ${q}$ we can expand Eq.\ \eqref{ReW+-} to obtain 
${\rm Re}(\omega^+) = -D {q}^{2}$, therefore the amplitude of perturbations decays exponentially for these modes. 
In the case of large wavelength perturbations the expansion to the lowest orders in ${q}$ yields
\begin{multline}
    {\rm Re}(\omega^+) =  \left[ \chi(1) -D \right]{q}^{2} \\
- \chi(1) {q}^{2} \left\{ \left[1-D+ \chi(1) \right] {q}^{2}  +\left(\mathbf{{V}} \cdot \mathbf{q}\right)^2 \right\}  +\mathcal{O} \left({q}^{6}\right).\label{ReW+}
\end{multline}
Thus, when $\chi(1)-D>0$ the positive branch has a band of unstable modes for small wave vectors. In the case of the standard KS model,  $\chi(1)=\chi_0 $, the above condition is equivalent to the aggregation threshold: $D/\chi_0<1$ \cite{keller:1970,nanjundiah:1973}. In Eq.\ \eqref{ReW+} we observe that the advection velocity appears at the order  $q^4$. This means that, when the above condition for instability is satisfied, there is always a band of unstable modes with long wavelengths around $q=0$, but the range of unstable modes and their growth rate decreases with the advection velocity.
This stabilizing effect of the flow 
is in contrast with the behavior of reaction-diffusion systems studied by Rovinsky and Menzinger
\cite{rovinsky:1992,rovinsky:1994} where the differential flow induces an instability at a finite wavelength. 
For the imaginary part of $\omega$, we have the following expansion of Eq.\ \eqref{ImW+-} for small wave vectors:
\begin{equation}
{\rm Im}(\omega^{+})=- \chi(1) q^2 \left(\mathbf{{V}} \cdot \mathbf{q}\right) + \mathcal{O} \left({q}^{5} \right). \label{ImW+} 
\end{equation}
Therefore, although the particle density is not directly advected by the flow, using Eq. \eqref{Vl} we see that the chemotaxis induces a phase velocity, $\mathbf{V^{\it l}}$, that is inversely proportional to the square of the wavelength and, consequently, it is small relative to the advection velocity.

\section{Pattern formation in one dimension} \label{1D}

\begin{figure}[!htmb]
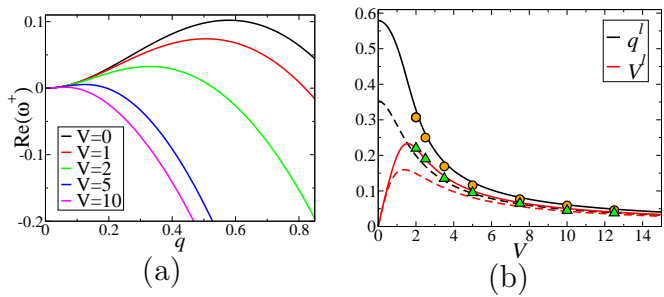

\begin{center}
\begin{minipage}[c]{0.475\linewidth}
\begin{center}
\includegraphics[width=\linewidth]{Figures/Fig1a.eps}
{\large (a)}
\end{center}
\end{minipage}\hspace*{ 0.05\linewidth}
\begin{minipage}[c]{0.475\linewidth}
 \begin{center}
\includegraphics[width=\linewidth]{Figures/Fig1b.eps}
{\large (b)}
\end{center}
\end{minipage}
\caption{(a) Real part of the dispersion relation ${\rm Re}(\omega^+)$ given by Eq.\ \eqref{ReW+-} as a function of $q$ for a 1D system with $\chi(u)=\chi_0 u(1-u/u_{max})$, for $D=1$, $\chi_0=2.5$, $u_{max}=4$ and different values of the advection velocity, $V$. (b) $q^{\it l}$ (black lines) and ${V^{\it l}}$ (red lines) as a function of the advection velocity. The solid lines represent the exact solution obtained from \eqref{ReW+-} and \eqref{ImW+-}. The dashed lines represent the approximate solution for small wavelengths. The velocity of the pattern and the dominant mode measured from the numerical simulations are represented by triangles and circles respectively. \label{Fig.1d}}
\end{center}
\end{figure}

Figure \ref{Fig.1d}(a) shows the effect of the advection velocity on the real part of the dispersion relation given by Eq.\ \eqref{ReW+-} for a $1D$ system. When  $V$ increases the wavenumber of the dominant mode corresponding to the maximum of $ {\rm Re}(\omega^+)$, $q^{\it l}$, decreases [Fig.\ \ref{Fig.1d}(b)]. 
In fact, from the expansion of Eq.\ \eqref{ReW+-} for small $q$ given by Eq.\ \eqref{ReW+} we have $q^{\it l}=\left[\nu / \left(2\mathcal{K}\right)\right]^{1/2}$ with $\nu= \chi(1) -D$ and $\mathcal{K}=  \chi(1) \left[(1-D)  + \chi(1)+V^2 \right]$. Thus for large $V$, the dominant wavelength in the linear regime, $\lambda^{\it l}=2\pi/q^{\it l}$, is proportional to the advection velocity. The phase velocity of the spatial pattern [Eq.\ \eqref{Vl}]  is also shown in Fig.\ \ref{Fig.1d}(b) as a function of the advection velocity for the exact dispersion relation [Eq.\ \eqref{ImW+-}]. We notice that the phase velocity has a maximum for a certain value of $V$. Thus, surprisingly, when the advection velocity is increased beyond this value, the phase velocity
of the pattern decreases as $V$ is increased. This non-monotonic dependence can also be shown using the expansion of ${\rm Im}(\omega^{+})$ for small wavenumber modes [Eq. \eqref{ImW+}] from where a compact analytical expression for the phase velocity is obtained [see Fig.\ \ref{Fig.1d}(b)]
\begin{equation}\label{1d_Vlb}
{V^{\it l}}=  \chi(1) V \left(q^{\it l}\right)^2 = \frac{  \left[ \chi(1)-D  \right]V} { 2 \left[ \left(1-D\right)+\chi(1)+{V}^{2} \right]}.
\end{equation}
Although this expression is only valid for small values of $q^{\it l}$, it is qualitatively similar to the exact solution and already shows that ${V^{\it l}}$ increases linearly for small values of $V$ and when the advection velocity exceeds a certain threshold, $V_t=\sqrt{\left(1-D \right) + \chi(1)}$, larger values of $V$ induces slower pattern movement.

It is interesting to discuss how the linear wavelength and the phase velocity are modified when $D$ is decreased, since in typical experiments $D_u/D_v\ll1$. For strong advection we can use the expansion of Eq.\ \eqref{ReW+-} for small $q$ to obtain the minimal value of the linear wavelength. Since in the aggregation regime, $0<D<\chi(1)$, the linear wavelength is an increasing function of $D$, for $D\rightarrow0$ we obtain $\lambda^{\it l}_{min}=2^{3/2}\pi \sqrt{1+\chi(1)+V^2}$, from where we see that the linear wavelength becomes larger when the chemosensitivity is increased. On the other hand, $V_t$ is a decreasing function of $D$. Thus, in the limit $D\rightarrow0$ the maximum phase velocity is reached when the advection is $V_t=\sqrt{1 + \chi(1)}$. Therefore, as we could expect, the efficiency of the particles following the chemical field depends on the chemosensitivity. For larger values of $ \chi(1)$ the cells can more accurately follow the chemical field up to larger values of the advection velocity.

For the numerical simulations we use the chemotactic response function: $\chi(u)=\chi_0 u (1-u/u_{max})$, that avoids the singularities associated with other models. Interestingly, we have also found that in the case of the standard linear response, $\chi(u)=\chi_0 u$, advection can suppress the singularity and produce qualitatively similar behavior to the previous function when $V$ is sufficiently large. 
We consider periodic boundary conditions with initial fields in the uniform steady state with a small amplitude random perturbation. The parameters in Eqs. \eqref{Eq.u3} and \eqref{Eq.v3} were set to $D=1$, $\chi_0=2.5$, and $u_{max}=4$ for the one- and two-dimensional simulations. 

\begin{figure}[!htmb]
\begin{center}
\begin{minipage}[c]{0.475\linewidth}
\begin{center}
\includegraphics[width=\linewidth]{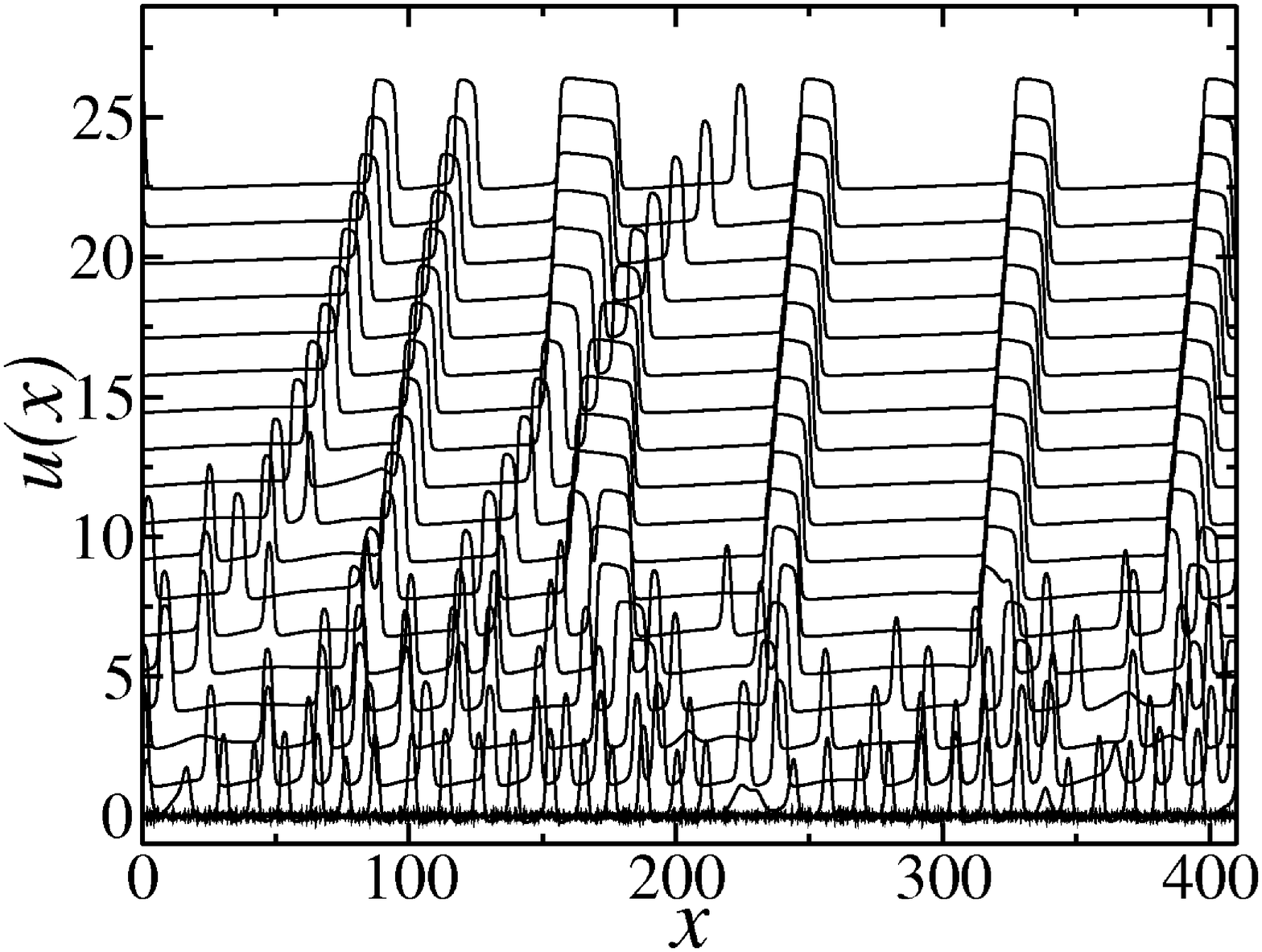}
{\large (a)}
\end{center}
\end{minipage}\hspace*{ 0.05\linewidth}
\begin{minipage}[c]{0.475\linewidth}
 \begin{center}
\includegraphics[width=\linewidth]{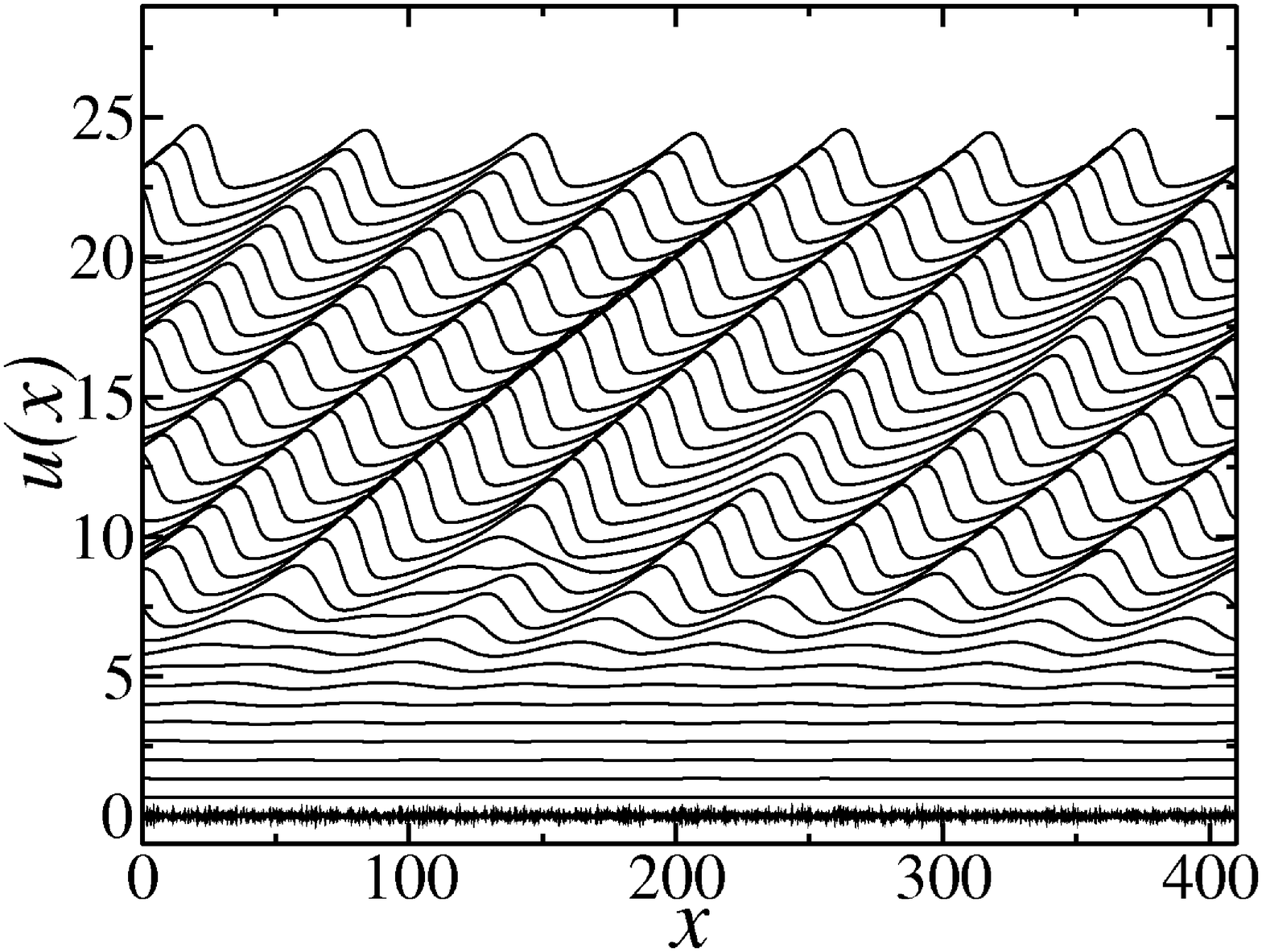}
{\large (b)}
\end{center}
\end{minipage}
\caption{(a) Particle density profiles at equally spaced times (profiles are offset vertically) between $t=0$ (bottom) and $t=3500$ (top) for: (a) $V=1$ and (b) $V=5$.\label{Fig.profiles}}
\end{center}
\end{figure}

\begin{figure}[!htmb]
\begin{center}
\begin{minipage}[c]{0.475\linewidth}
\begin{center}
\includegraphics[width=\linewidth]{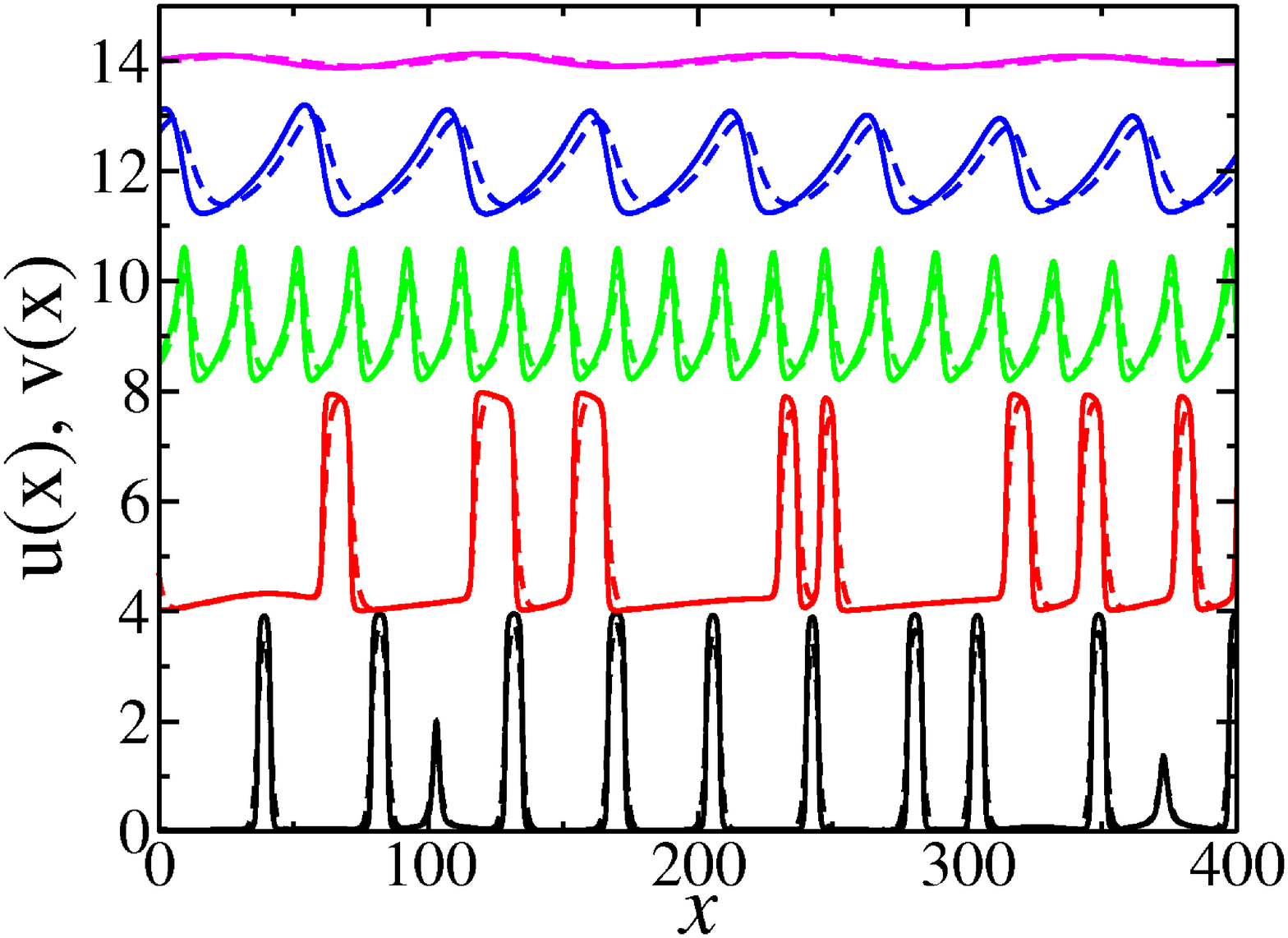}
{\large (a)}
\end{center}
\end{minipage}\hspace*{ 0.05\linewidth}
\begin{minipage}[c]{0.475\linewidth}
 \begin{center}
\includegraphics[width=\linewidth]{Figures/Fig3b.eps}
{\large (b)}
\end{center}
\end{minipage}
\caption{(a) Density, $u$, (solid line) and signal concentration, $v$, (dashed line) profiles at $t=3500$ for $V=0,1,2,5,$ and $10$, from bottom to top respectively (profiles are offset vertically). (b) Temporal evolution of the lateral pattern wavelength, $\lambda(t)$, for different values of $V$. Error bars represent the standard deviation calculated from 150 different random initial conditions.
\label{Fig.u_l}}
\end{center}
\end{figure}

Representative one-dimensional numerical simulations of the particle density profiles $u(x,t)$ are shown in Fig.\ \ref{Fig.profiles} for two different values of the advection velocity. In the case of $V=1$ [Fig.\ \ref{Fig.profiles}(a)] we observe a slow coarsening process which tends to reduce the number of structures as observed in the system without advection \cite{hillen:2001}.
However, for $V=5$ [Fig.\ \ref{Fig.profiles}(b)] after a short transient a periodic pattern with constant wavelength develops.
As shown in Fig.\ \ref{Fig.profiles}, in the presence of advection the $x\rightarrow -x$ symmetry of the profiles is broken [see also Fig.\ \ref{Fig.u_l}(a)] and the pattern propagates in the positive direction (i.e. in the direction of the advection velocity). In the absence of coarsening (i.e. for values of $V\gtrsim2$) the pattern moves uniformly with a well defined velocity as shown in Fig.\ \ref{Fig.profiles}(b). The velocity of the pattern was measured for different values of $V$ and is plotted in Fig.\ \ref{Fig.1d}(b). The measured velocity agrees well with the analytical results obtained from the linear stability analysis. Independently of the initial chemical concentration, after a transient time the behavior of the chemoattractant $v$ is very similar to the density profiles, as shown in Fig.\ \ref{Fig.u_l}(a) where the profiles of $u$ and $v$ are plotted together for different values of $V$. For large advection velocities the particles can not ``follow'' the chemical gradients, the two profiles are more different and the aggregates are more spread out. Figure \ref{Fig.u_l}(b) shows the temporal evolution of the wavelength $\lambda(t)$ (measured as two times the average distance between consecutive minima and maxima) for different values of $V$. Aggregation starts earlier for larger values of $V$ and for strong enough advection values (larger than $V_t$) coarsening is interrupted and the wavelength of the pattern becomes constant with a final value which is proportional to $V$ as predicted by the linear analysis [see Fig.\ \ref{Fig.1d}(b)]. Thus the nonlinear effects (such as coarsening) are avoided for large values of $V$ and the system remains in the linear regime. Similar results are obtained for the standard KS model, $\chi(u)=\chi_0 u$, with the difference that for slow advection the numerical simulations do not reach a stationary state, indicating an aggregation singularity with unlimited growth of the density in some points.


\section{Pattern formation in two dimensions} \label{2D}

\begin{figure}[!htmb]
\begin{center}
\begin{minipage}[c]{0.475\linewidth}
\begin{center}
\includegraphics[width=\linewidth]{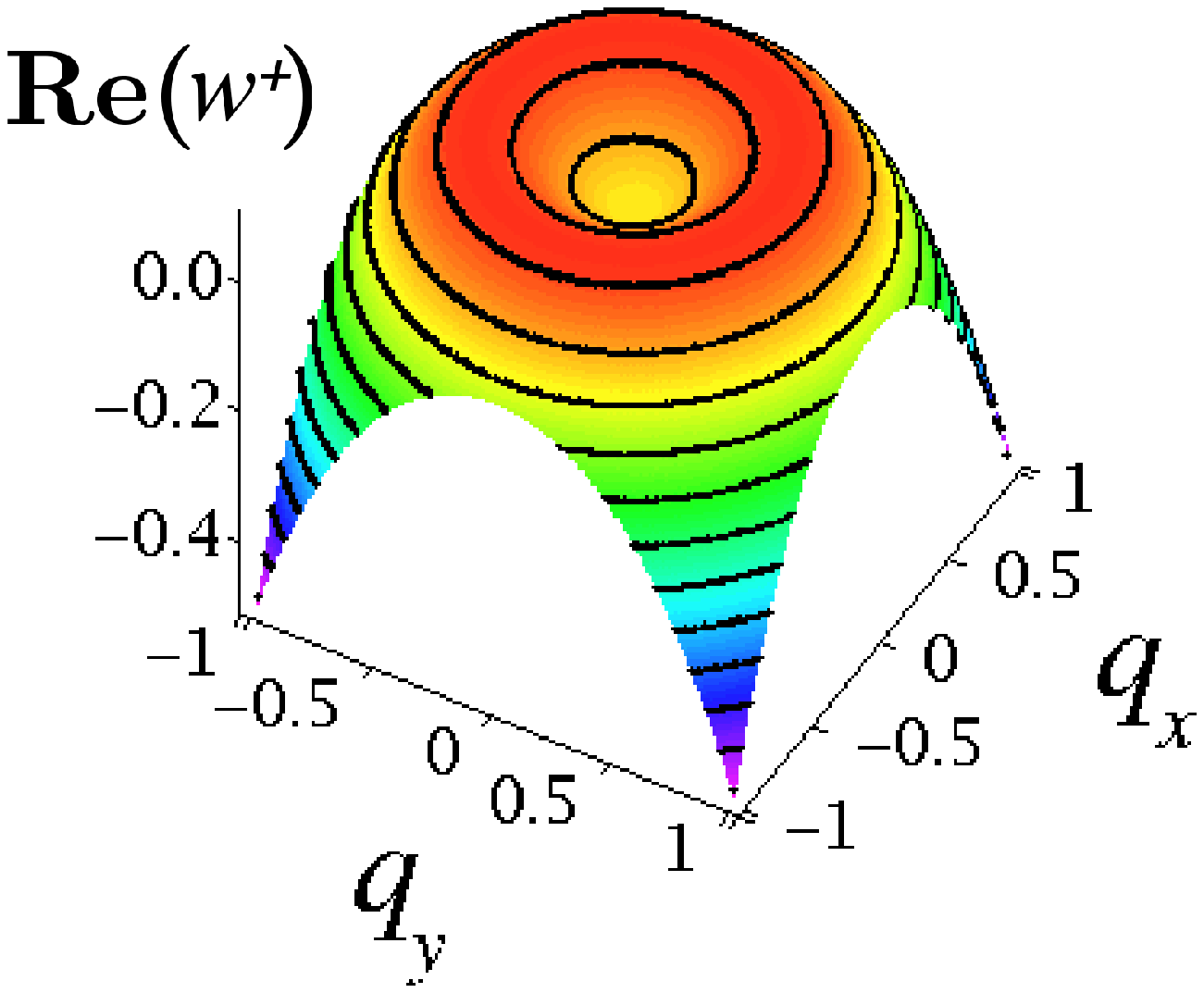}
{\large (a)}
\end{center}
\end{minipage}
\begin{minipage}[c]{0.475\linewidth}
\begin{center}
\includegraphics[width=\linewidth]{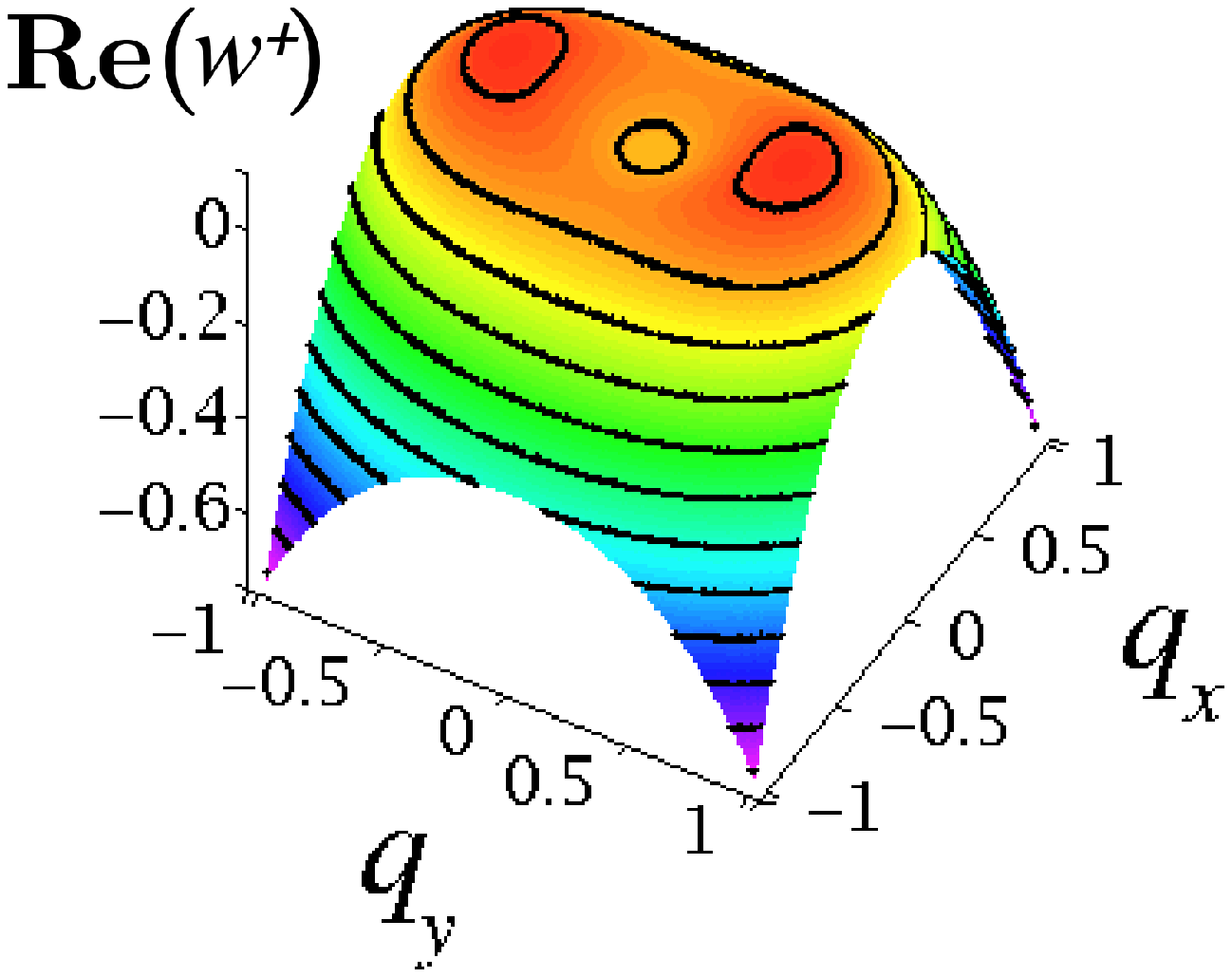}
{\large (b)}
\end{center}
\end{minipage}
\caption{Real part of the dispersion relation ${\rm Re}(\omega^+)$ given by Eq.\ \eqref{ReW+-} as a function of $q_x$ and $q_y$ for a $2D$ system with (a) $V=0$ and (b) $V=2$. \label{Fig.2d}}
\end{center}
\end{figure}

The effect of the advection term in the real part of the dispersion relation given by Eq.\ \eqref{ReW+-} for a $2D$ system is shown in Fig.\ \ref{Fig.2d}. Without loss of generality we assume that the advection is along the $x$-axis. Thus, we can write 
$\mathbf{{V}}=(V,0)$ and we plot ${\rm Re}(\omega^+)$ as a function of $q_x$ and $q_y$ for different values of ${V}$. When $V=0$ the system is isotropic and all the wavevectors within a distance $q^{\it l}$ from the origin maximize ${\rm Re}(\omega^+)$ [Fig.\ \ref{Fig.2d}(a)]. When ${V}$ is nonzero, the maximum is oriented along the $y$-axis but the absolute value is not affected by the value of ${V}$ [Fig.\ \ref{Fig.2d}(b)]. The  effect of the advection on the wavelength and orientation of the pattern can be analyzed in the small $q$ limit using Eq.\ \eqref{ReW+}. It can be shown that the dominant mode is always oriented along the $y$-axis and its value is $q_y^{\it l}=\left[\nu / \left(2\mathcal{K}_{y}\right)\right]^{1/2}$ where $\mathcal{K}_{y}=  \chi(1) \left[(1-D) + \chi(1)\right]$ and $\nu$ is the same as defined for the $1D$ case (see Appendix). Thus, in contrast to the one-dimensional system, the observed dominant wavelength of the pattern in the linear regime is independent of the advection velocity $V$, although the features of the pattern are altered by the symmetry breaking induced by the advection. For the phase velocity, using the expansion of ${\rm Im}(\omega^{+})$ for small $q$ and Eq.\ \eqref{Vl} we have $\mathbf{V^{\it l}} \cdot \mathbf{q^{\it l}}/{q^{\it l}}=  \chi(1) q^{\it l} \left(\mathbf{V}\cdot \mathbf{q^{\it l}}\right)=0$ because $\mathbf{V}=(V,0)$ and $\mathbf{q^{\it l}}=(0,q_y^{\it l})$. Therefore, the dominant mode in the pattern is not moving since it is oriented in the $y$ axis that is perpendicular to the advection velocity. However, for modes in the $x$-axis, $q_x$, the phase velocity in this direction is ${V^{\it l}_x}= {\chi(1)} q_x^2 {V}$ which is proportional to the advection velocity and increases with the wavenumber.

\begin{figure}[!htmb]
\begin{center}
\begin{minipage}[c]{0.2\linewidth}
\begin{center}
\includegraphics[width=\linewidth]{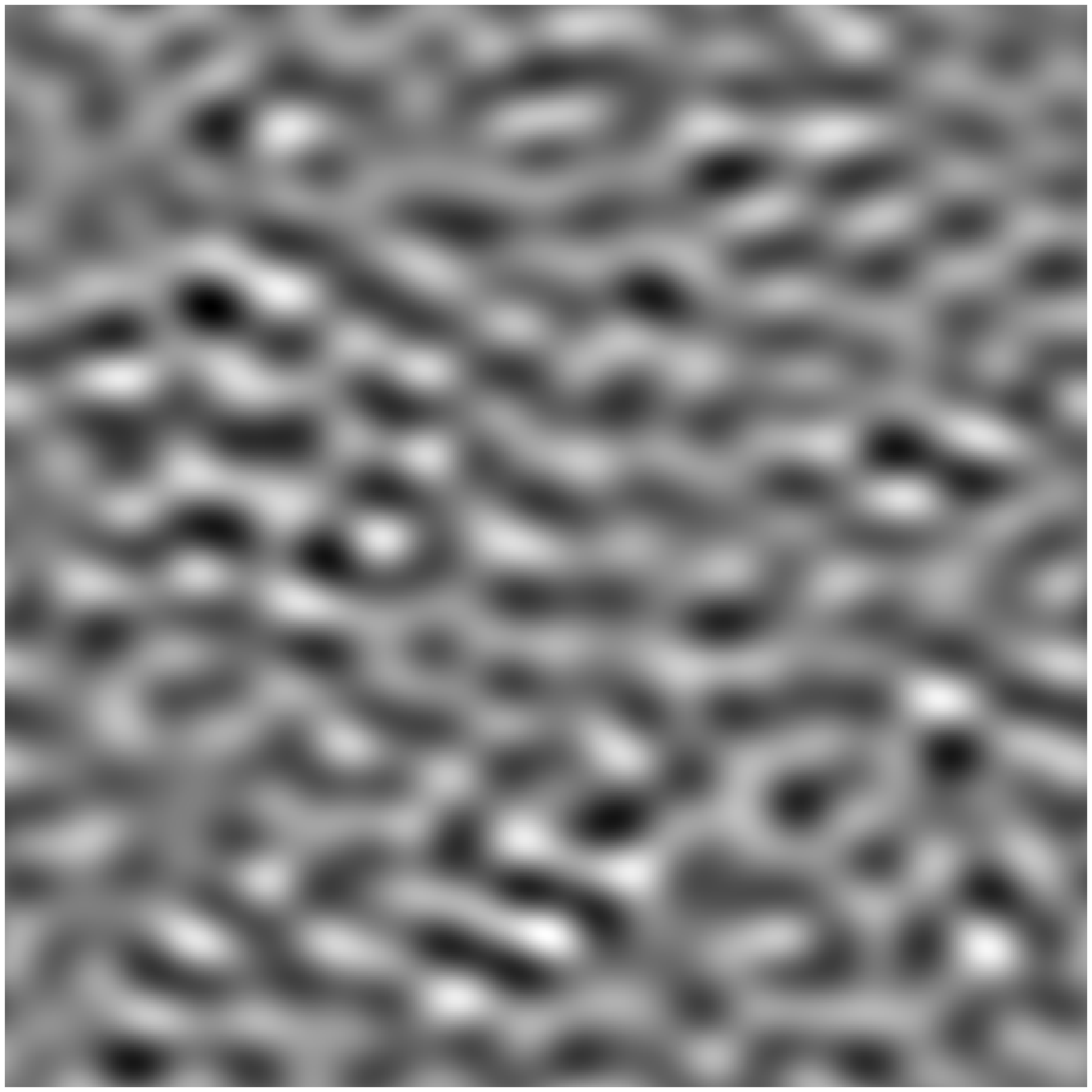}
\end{center}
\end{minipage}\hspace*{ 0.01\linewidth}
\begin{minipage}[c]{0.2\linewidth}
 \begin{center}
\includegraphics[width=\linewidth]{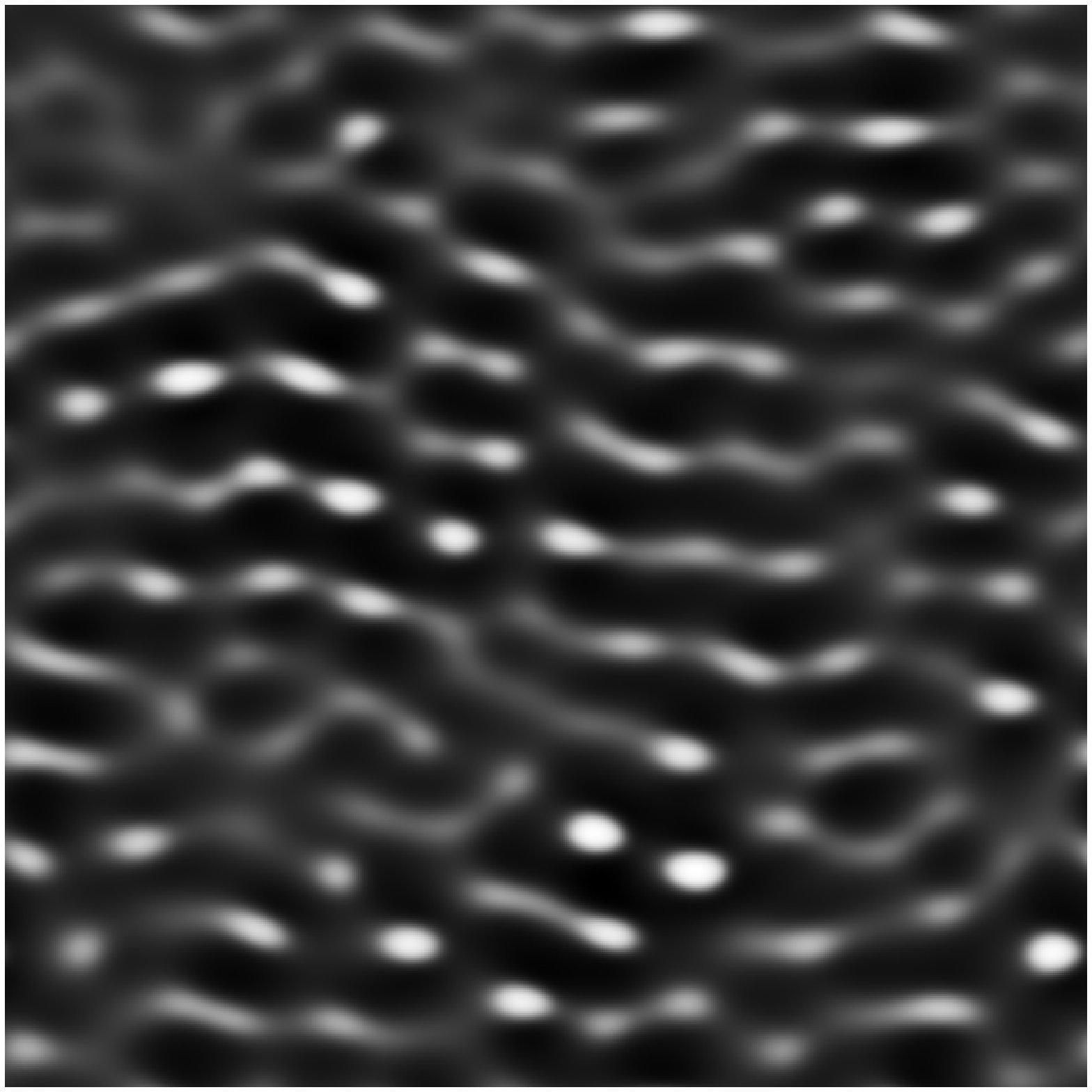}
\end{center}
\end{minipage}\hspace*{ 0.01\linewidth}
\begin{minipage}[c]{0.2\linewidth}
\begin{center}
\includegraphics[width=\linewidth]{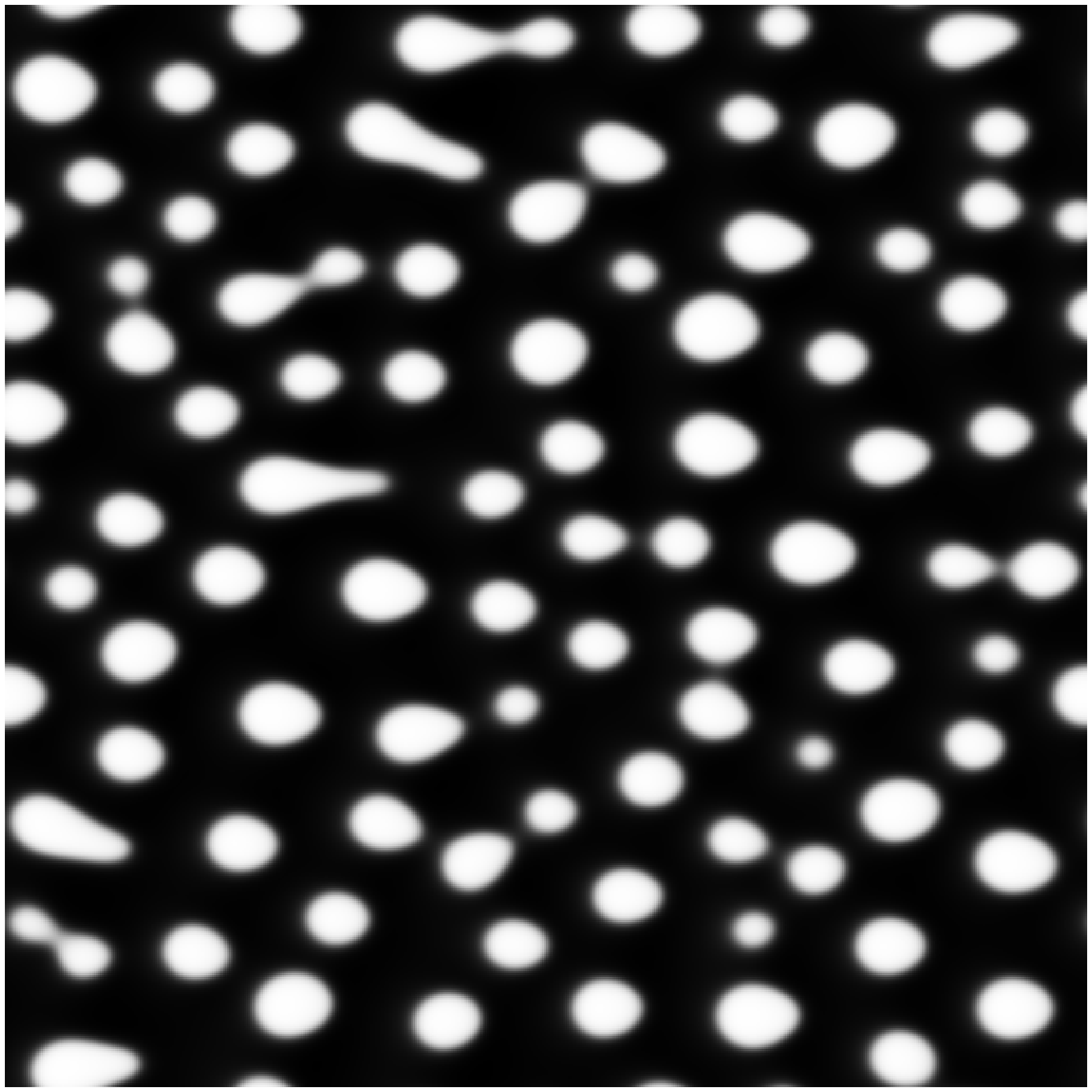}
\end{center}
\end{minipage}\hspace*{ 0.01\linewidth}
\begin{minipage}[c]{0.2\linewidth}
\begin{center}
\includegraphics[width=\linewidth]{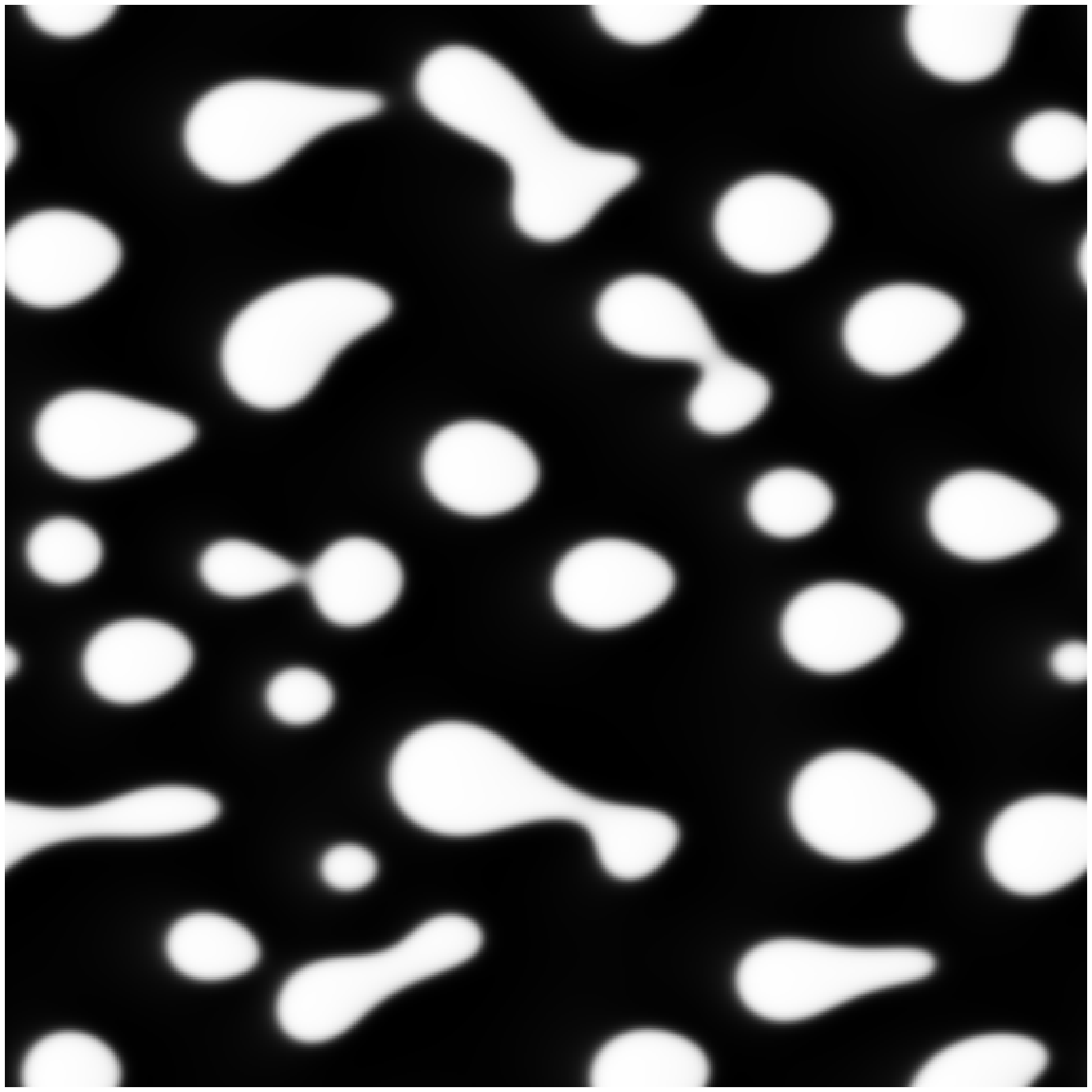}
\end{center}
\end{minipage}
\vspace{0.03\linewidth}
\\
\begin{minipage}[c]{0.2\linewidth}
\begin{center}
\includegraphics[width=\linewidth]{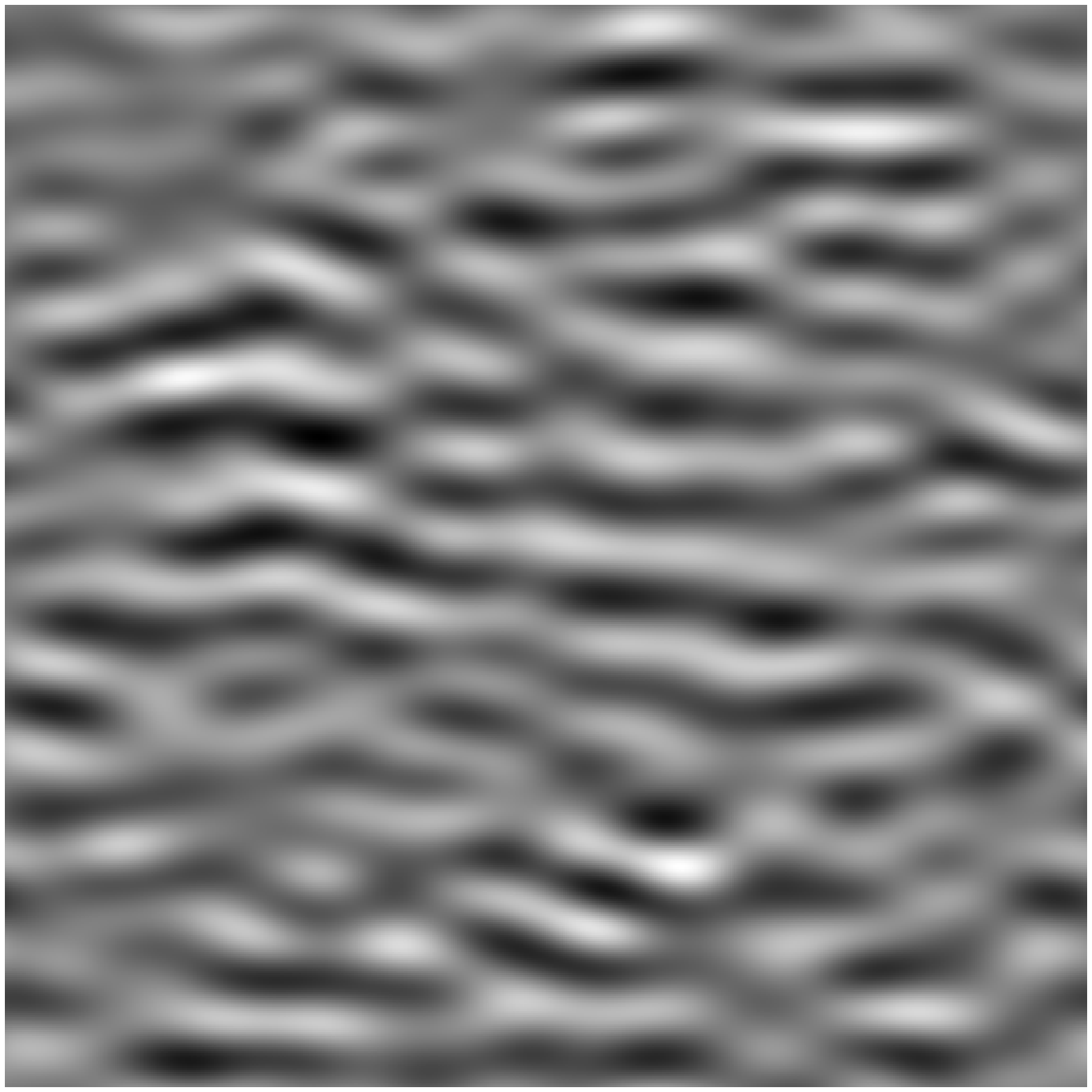}
\end{center}
\end{minipage}\hspace*{ 0.01\linewidth}
\begin{minipage}[c]{0.2\linewidth}
 \begin{center}
\includegraphics[width=\linewidth]{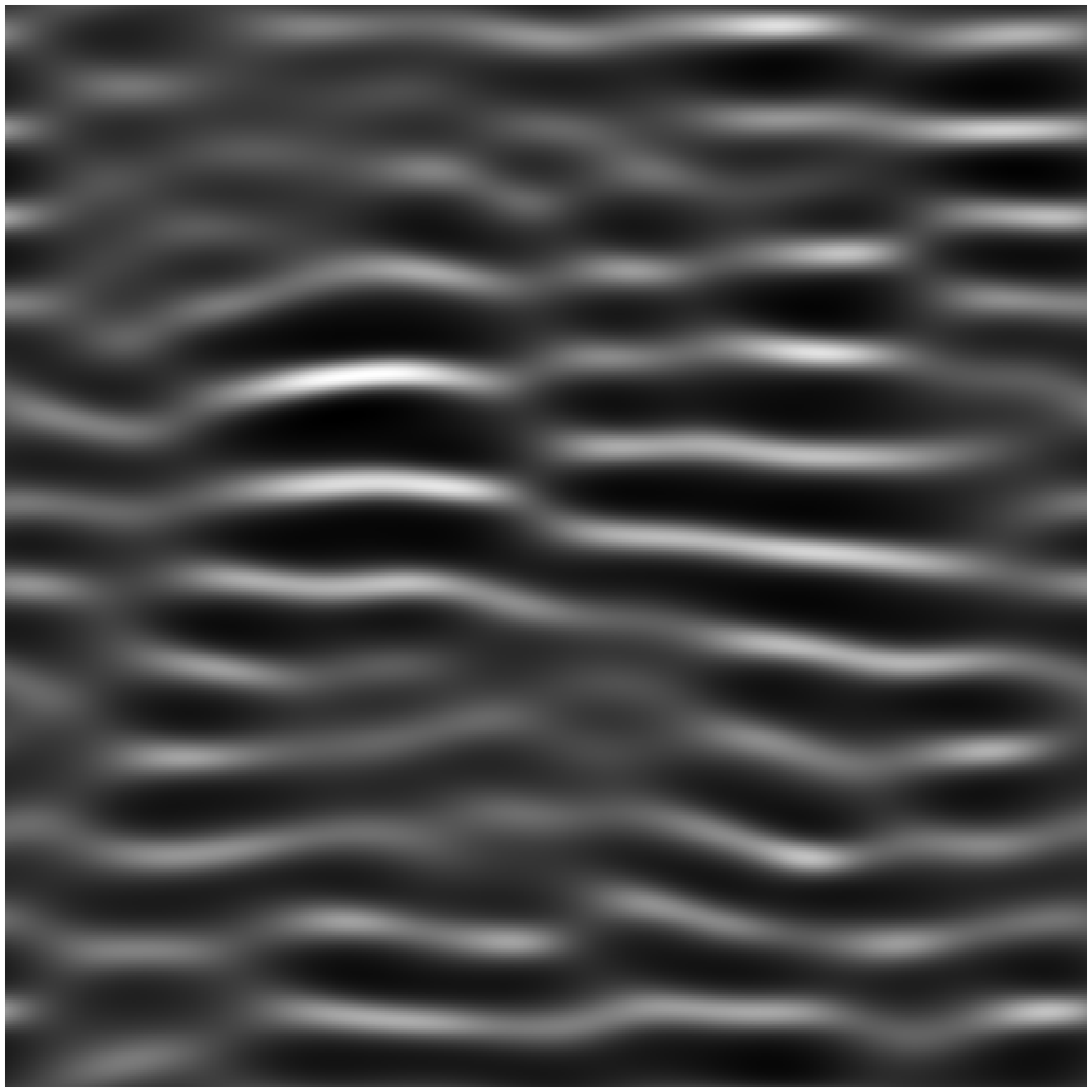}
\end{center}
\end{minipage}\hspace*{ 0.01\linewidth}
\begin{minipage}[c]{0.2\linewidth}
\begin{center}
\includegraphics[width=\linewidth]{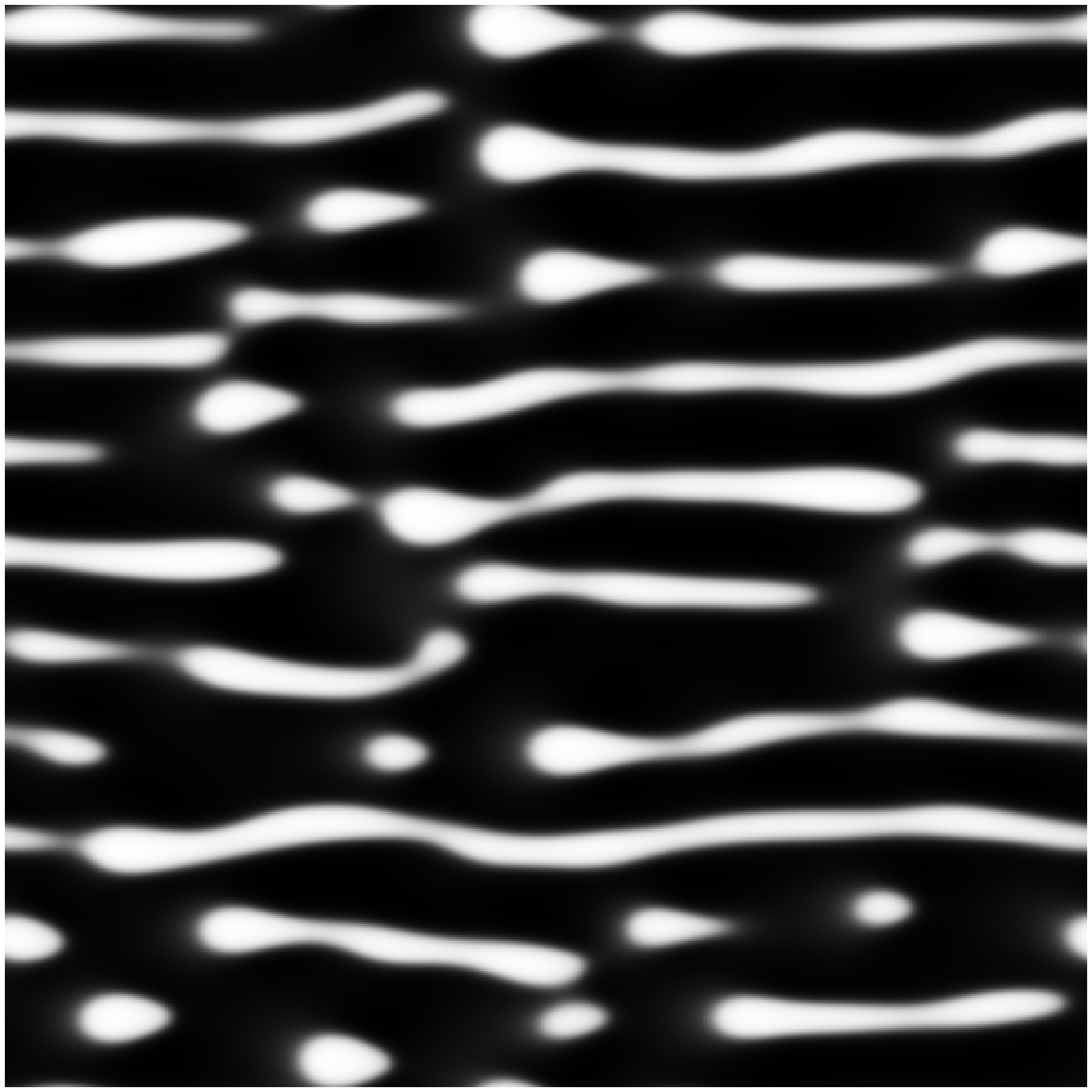}
\end{center}
\end{minipage}\hspace*{ 0.01\linewidth}
\begin{minipage}[c]{0.2\linewidth}
\begin{center}
\includegraphics[width=\linewidth]{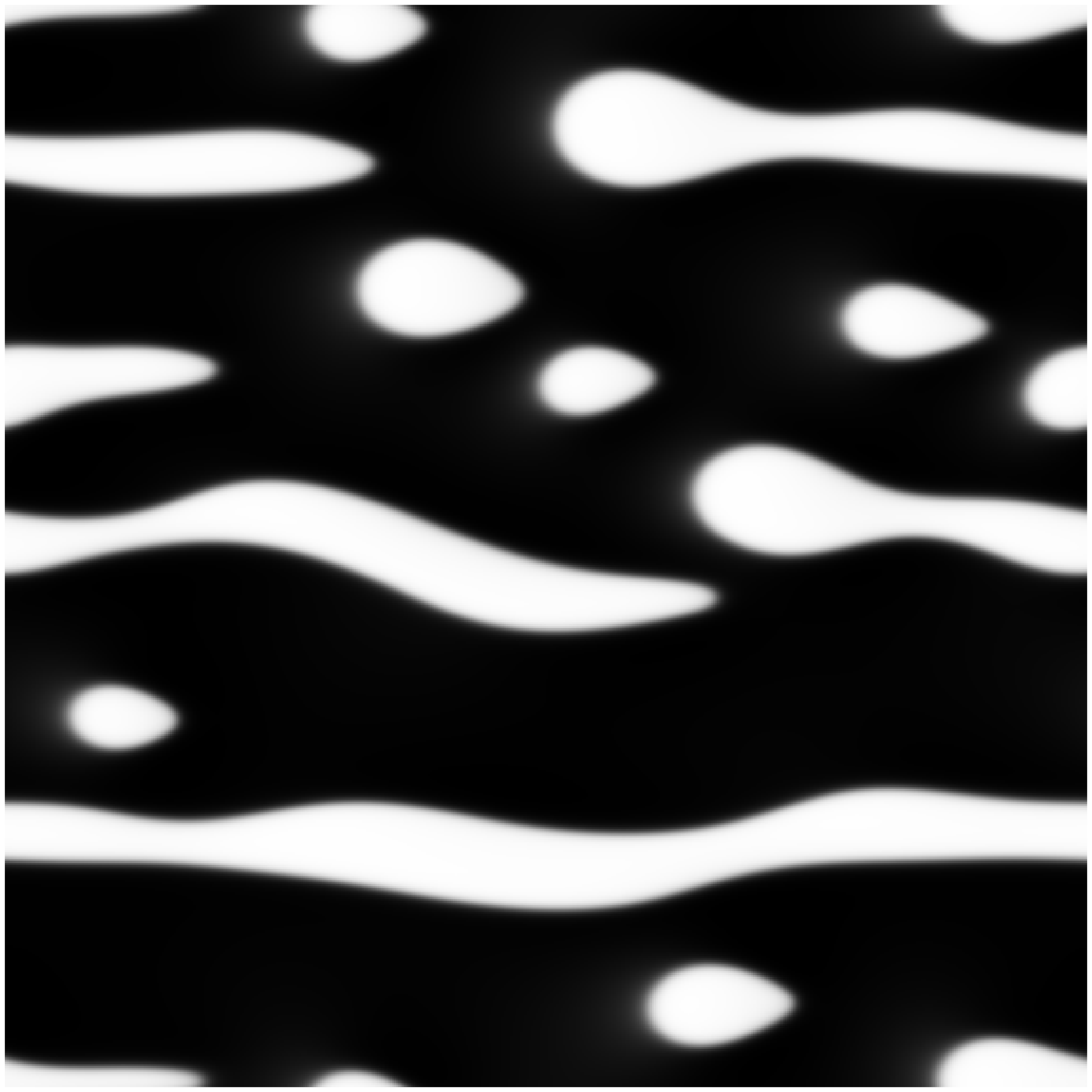}
\end{center}
\end{minipage}
\vspace{0.03\linewidth}
\\
\begin{minipage}[c]{0.2\linewidth}
\begin{center}
\includegraphics[width=\linewidth]{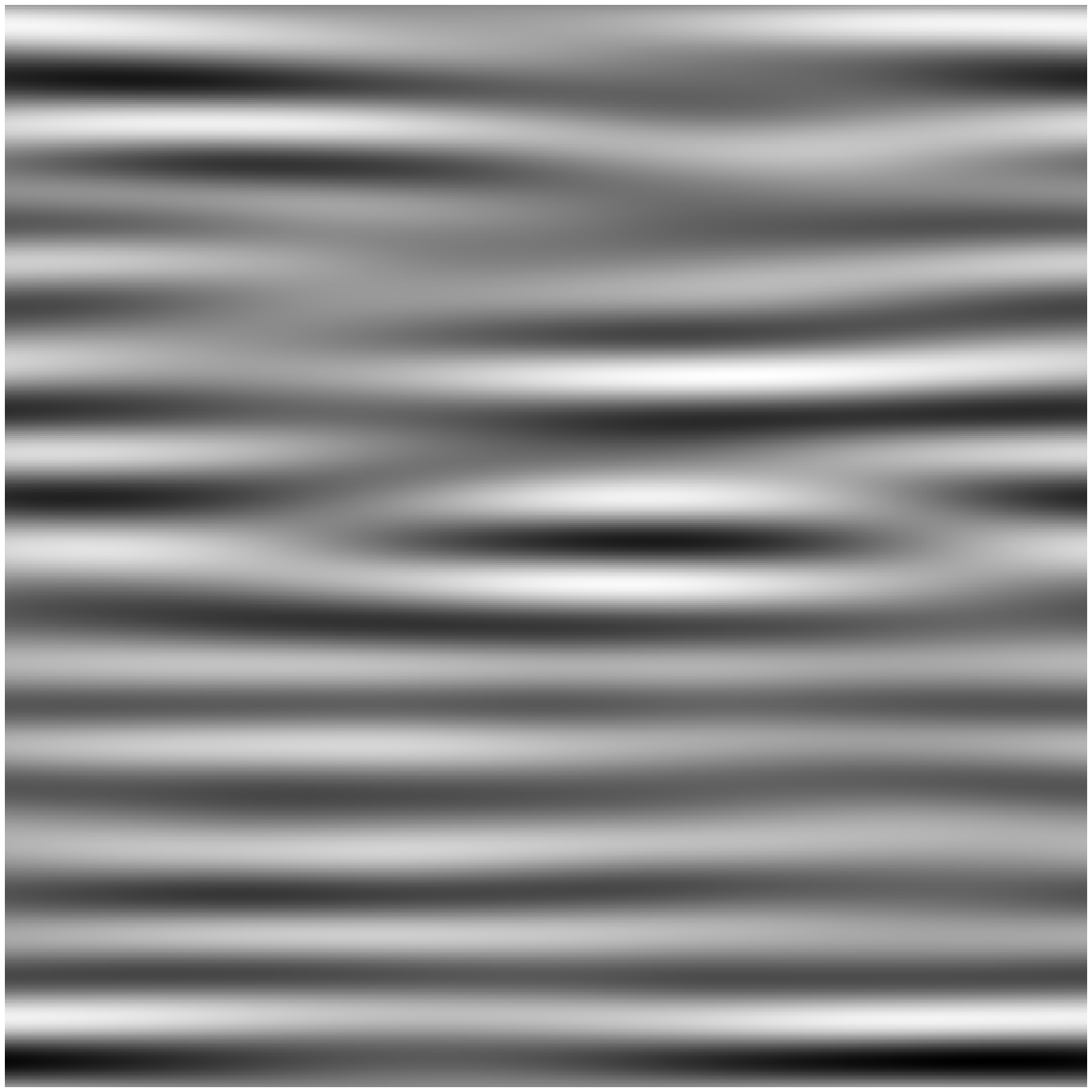}
\end{center}
\end{minipage}\hspace*{ 0.01\linewidth}
\begin{minipage}[c]{0.2\linewidth}
 \begin{center}
\includegraphics[width=\linewidth]{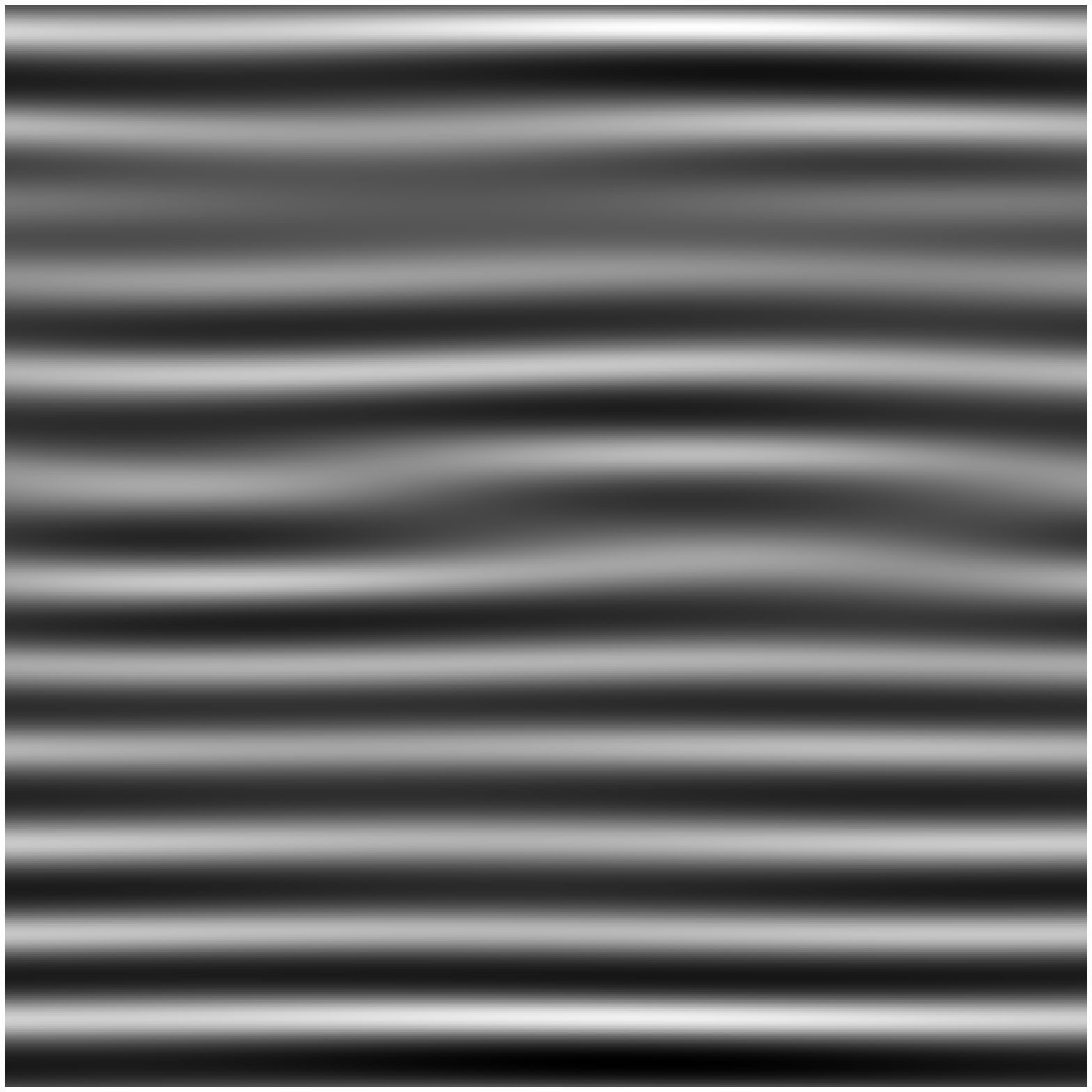}
\end{center}
\end{minipage}\hspace*{ 0.01\linewidth}
\begin{minipage}[c]{0.2\linewidth}
\begin{center}
\includegraphics[width=\linewidth]{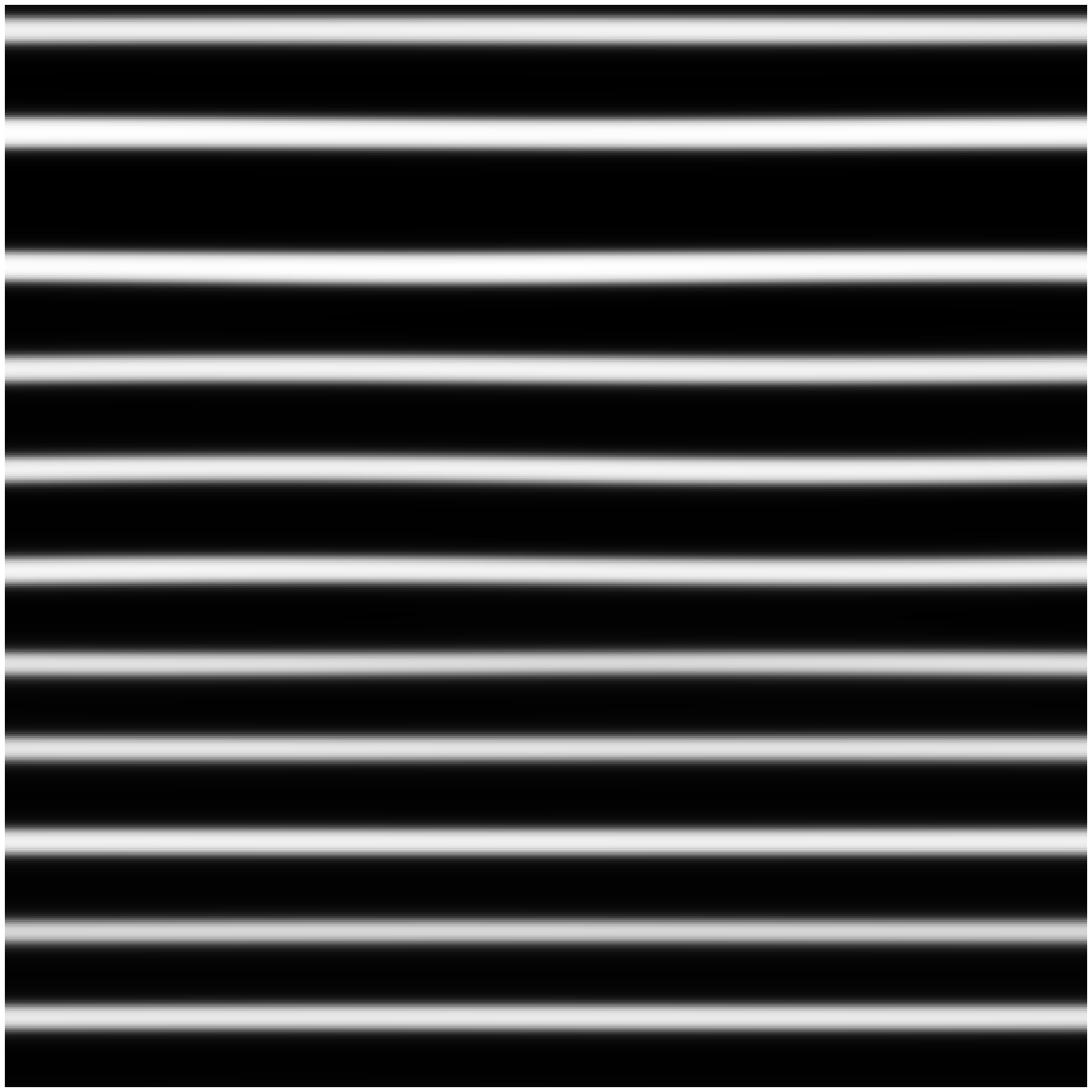}
\end{center}
\end{minipage}\hspace*{ 0.01\linewidth}
\begin{minipage}[c]{0.2\linewidth}
\begin{center}
\includegraphics[width=\linewidth]{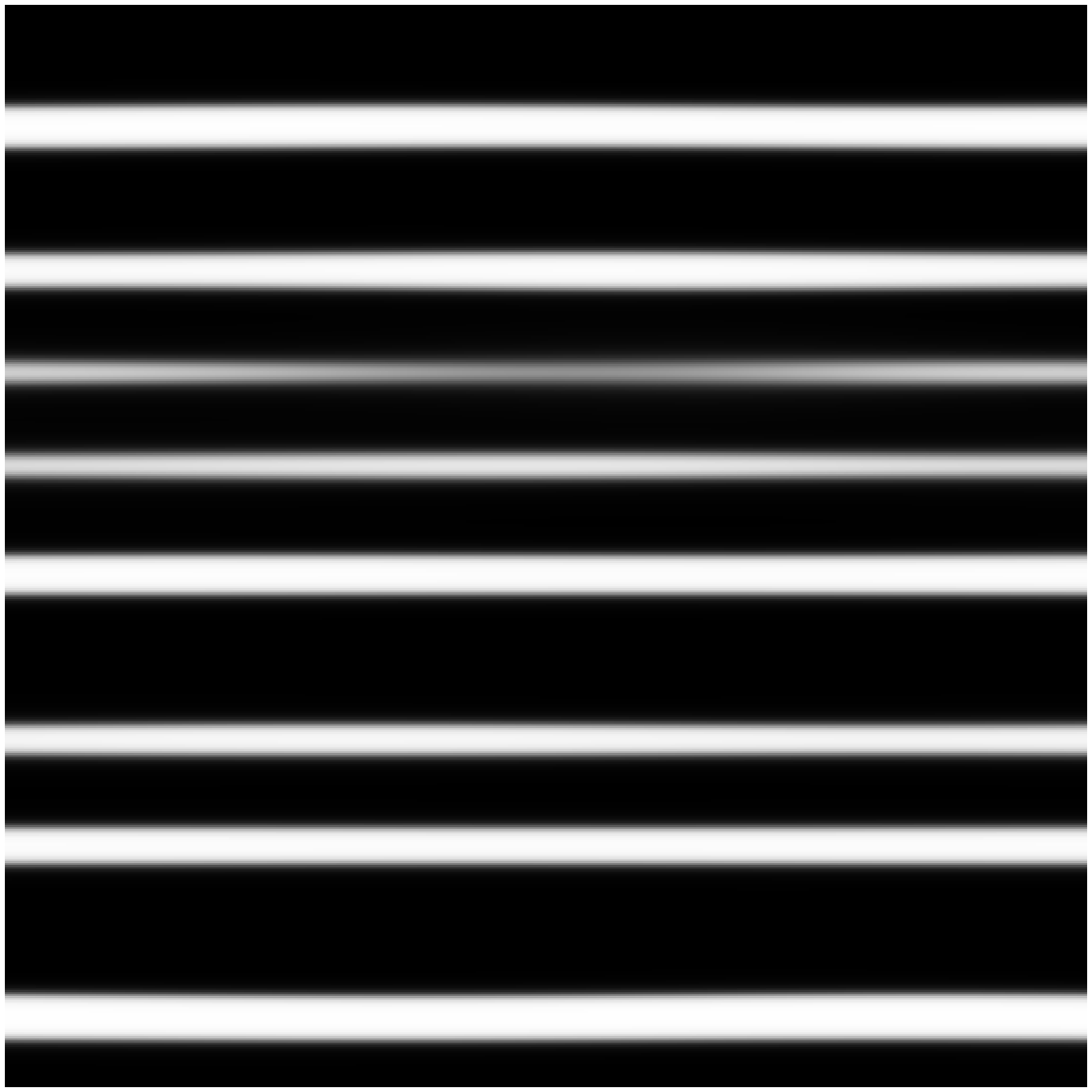}
\end{center}
\end{minipage}
\caption{Temporal evolution of $u$ for different values of the advection velocity: $V=1$ (first row), $V=2$ (second row), $V=10$ (third row); and different times: $t=35$ (first column), $t=70$ (second column), $t=200$ (third column), $t=870$ (fourth column). The $x$ axis is horizontally oriented and the system size is $128 \times 128$. See also supplementary movies \cite{EPAPS}.\label{Fig.2D}}
\end{center}
\end{figure}

Two-dimensional numerical simulations were performed using 
a cubic interpolation semi-lagrangian scheme \cite{durran} for the advection. For the spatial discretization of the particle density we used the method proposed recently by Grima and Newman \cite{grima:2004}, that allows for an accurate analysis of the evolution of the system without the dissipative effects of other schemes. 
In the absence of advection a slow coarsening process occurs in which, after long times, the organisms accumulate into a single aggregate \cite{hillen:2001}. When advection takes place the resulting pattern is not isotropic and the dominant wave vector is oriented along the $y$ axis. The temporal evolution of $u$ for different values of the advection velocity is shown in Fig.\ \ref{Fig.2D}. As in the $1D$ case, the evolution of the chemical field is very similar to the density field (see supplementary movies \cite{EPAPS}). As predicted above, the pattern does not propagate in the $y$ direction and the coarsening process along the $y$ axis is not affected by the advection (see supplementary movies \cite{EPAPS}). 
Thus, in this direction the pattern features remain the same when the advection velocity is increased. A different scenario is found in the $x$ direction where the pattern moves with a velocity which increases with $V$. The simulations also show that smaller aggregates move faster than larger ones, consistently with the dependence of the phase velocity on the wave-number in the linear analysis. As in the $1D$ case, the characteristic wavelength of the pattern in the $x$ direction increases with $V$. For slow advection there is a clear coarsening which leads to smaller propagation velocity of the aggregate as predicted above. For finite system sizes and large values of the advection velocity, the wavelength of the dominant linear instability in the $x$ direction becomes larger than the system size and the resulting pattern is homogeneous in the $x$ direction, observing a static pattern in this direction (see Fig.\ \ref{Fig.2D} for $V=10$).

\section{Conclusions} \label{Conclusions}


In this work we have studied theoretically how an advected uniform flow influences the aggregation dynamics in Keller-Segel type models. We found that, in the presence of a differential flow, an advective instability produces a pattern moving in the direction of the flow. Interestingly, although the organisms are not directly advected, the advection of the chemotactic signal induces a movement of the particle density as the organisms try to follow regions of high chemical concentrations. When the organisms mobility due to chemotaxis is weak in comparison to the advective flow (which is typically much faster), the balance between the chemotactic flux of the organism density and the advective transport of the chemical field breaks down and it is then restored by a change in the characteristic of the spatial patterns, that become strongly anisotropic with elongated stripe-like structures aligned to the direction of the flow. Furthermore, as shown above, for flows larger than a certain threshold the organism can not follow the chemotaxis signal reducing their velocity and eventually preventing the formation of large aggregates. Thus, the presence of a linear advection term inhibits the formation of large gradients and diminish the nonlinear effects. This stabilizing effect may completely stop the chemotactic aggregation and coarsening process in the direction of the flow. Although a simple uni-directional flow can not suppress the coarsening in the perpendicular direction, preliminary results with more general non-uniform time-dependent velocity fields in $2D$ show that aggregation may be halted preventing the appearance of singularities associated to the KS models. This is similar to the arrested coarsening process observed in binary mixtures \cite{berti:2005,naraigh:2007}.

The theoretical results on the distribution of chemotactic cell populations under the influence of a differential
flow are relevant for various natural and artificial systems including biofilms and could also be studied
experimentally in the context of Dictyostelium aggregation. This work may also provide a starting point for
the study of more general biological pattern formation phenomena in advected environments.

\appendix
\section{Linear pattern orientation} \label{app.A}

The experimentally observed pattern is mainly oriented along the direction which
yields the maximum value of the real part of the dispersion relation and its
wavelength is associated to the wave vector $\mathbf{q^{\it l}}=(q_x^{\it l}, q_y^{\it l})$. This vector verifies 
\begin{equation}\label{min}
\left[ \frac {\partial  {\rm Re}(\omega^+)}{\partial q_x}\right]_{\mathbf{q}^{\it l}}= \left[\frac {\partial  {\rm Re}(\omega^+)}{\partial q_y}\right]_{\mathbf{q}^{\it l}}=0, 
\end{equation}
which have the following real independent solutions
\begin{eqnarray}\label{kas}
    \mathbf{q_0}&=\left(0,0\right),\,
    \mathbf{q_1}=\left(\sqrt \frac{{} \nu}{2\mathcal{K}} ,0\right),\,
    \mathbf{q_2}=\left(0, \sqrt \frac{{} \nu}{2\mathcal{K}_{y}}\right),
\end{eqnarray}
where $\nu$, $\mathcal{K}$, and $\mathcal{K}_{y}$ are defined as in the main text and assumed positive. 
In order to decide which of the remaining solutions provide the absolute
maximum of ${\rm Re}(\omega^+)$, we finally substitute the wave
vectors given by \eqref{kas} into Eq.\ \eqref{ReW+}; we obtain simply
\begin{equation}
\left[{\rm Re}(\omega^+)\right]_{\mathbf{q_0}}=0,\, \left[{\rm Re}(\omega^+)\right]_{\mathbf{q_1}}=\frac{ \nu^2}{4\mathcal{K}},\,
\left[{\rm Re}(\omega^+)\right]_{\mathbf{q_2}}=\frac{ \nu^2}{4\mathcal{K}_{y}}, 
\end{equation}
from where, since $\mathcal{K} > \mathcal{K}_{y}$ for $V\neq0$, we conclude that $\mathbf{q}^{\it l}=\mathbf{q_2}$.

\begin{acknowledgments}
This work was supported by Science Foundation Ireland RFP research grant and a Centre for Science Engineering and Technology grant for SBI. Computational facilities were provided
by ICHEC.
\end{acknowledgments}


\bibliography{bib}

\end{document}